\documentclass[final,english]{bullsrsl}[2022/06/15]

\usepackage[latin1]{inputenc}
\usepackage[T1]{fontenc}

\usepackage{natbib} 
\usepackage{graphicx}

\begin{document}
\title{Detection and Identification of Asteroids with the 4-m ILMT
}

%\author[affil={1}]{Kuntal}{Misra}
\author[affil={1}, corresponding]{Anna}{Pospieszalska-Surdej}
\author[affil={2,3}]{Bhavya}{Ailawadhi}
\author[affil={4,5}]{Talat}{Akhunov}
\author[affil={6}]{Ermanno}{Borra}
\author[affil={2,7}]{Monalisa}{Dubey}
\author[affil={2,7}]{Naveen}{Dukiya}
\author[affil={8}]{Jiuyang}{Fu}
\author[affil={8}]{Baldeep}{Grewal}
\author[affil={8}]{Paul}{Hickson}
\author[affil={2}]{Brajesh}{Kumar}
\author[affil={2}]{Kuntal}{Misra}
\author[affil={2,3}]{Vibhore}{Negi}
\author[affil={2,9}]{Kumar}{Pranshu}
\author[affil={8}]{Ethen}{Sun}
\author[affil={1}]{Jean}{Surdej}
%\author[affil={9,10}]{Jean}{Surdej}
\affiliation[1]{Institute of Astrophysics and Geophysics, University of Li\`{e}ge, All\'{e}e du 6 Ao$\hat{\rm u}$t 19c, 4000 Li\`{e}ge, Belgium}
\affiliation[2]{Aryabhatta Research Institute of observational sciencES (ARIES), Manora Peak, Nainital, 263001, India}
\affiliation[3]{Department of Physics, Deen Dayal Upadhyaya Gorakhpur University, Gorakhpur, 273009, India}
\affiliation[4]{National University of Uzbekistan, Department of Astronomy and Astrophysics, 100174 Tashkent, Uzbekistan}
\affiliation[5]{ Ulugh Beg Astronomical Institute of the Uzbek Academy of Sciences, Astronomicheskaya 33, 100052 Tashkent, Uzbekistan}
\affiliation[6]{Department of Physics, Universit\'{e} Laval, 2325, rue de l'Universit\'{e}, Qu\'{e}bec, G1V 0A6, Canada}
\affiliation[7]{Department of Applied Physics, Mahatma Jyotiba Phule Rohilkhand University, Bareilly, 243006, India}
\affiliation[8]{Department of Physics and Astronomy, University of British Columbia, 6224 Agricultural Road, Vancouver, BC V6T 1Z1, Canada}
\affiliation[9]{Department of Applied Optics and Photonics, University of Calcutta, Kolkata, 700106, India}

%\affiliation[10]{Astronomical Observatory Institute, Faculty of Physics, Adam Mickiewicz University, ul. Sloneczna 36, 60-286 Poznan, Poland}

\correspondance{anna.pospieszalska@uliege.be}
\date{6th May 2023}
\maketitle

% \author[affil1]{FirstName (+ MiddleInitials if necessary)}{FamilyName}
% \author[affil2]{...}{}
% \equalcontribauthor[]{}{} % Maximum two --> counter
% \consortium[affil]{Consortium Name}
% With consortium: affiliation will be set to "See Appendix 1 for a full
% list of consortium members and their respective affiliations
% \affiliation[affil1]{...}
% \affiliationq[affil2]{...}

% \correspondence[]{}
% No explicit corresponding author: use first author
% 

% Abstract of the paper in the same language as the paper
\begin{abstract}
A very unique strength of the Devasthal Observatory is its capability of detecting optical transients with the 4-m International Liquid Mirror Telescope (ILMT) and to rapidly follow them up using the 1.3-m Devasthal Fast Optical Telescope (DFOT) and/or the 3.6-m Devasthal Optical Telescope (DOT), installed right next to it. 
In this context, we have inspected 20 fields observed during 9 consecutive nights in October-November 2022 during the first commissioning phase of the ILMT. Each of these fields has an angular extent of $22^\prime$ in declination by $9 \times 22^\prime$ in right ascension. Combining both a visual search for optical transients and an automatic search for these using an image subtraction technique (see the ILMT poster paper by Pranshu et al.), we report a total of 232 significant transient candidates. After consulting the Minor Planet Center database of asteroids, we could identify among these 219 positions of known asteroids brighter than $V=22$. These correspond to the confirmed positions of 78 distinct known asteroids. Analysis of the remaining CCD frames covering 19 more fields (out of 20) should lead to an impressive number of asteroids observed in only 9 nights.
The conclusion is that in order to detect and characterize new supernovae, micro-lensing events, highly variable stars, multiply imaged quasars, etc. among the ILMT optical transients, we shall first have to identify all known and new asteroids. 
Thanks to its large diameter and short focal length (f/D $\sim$ 2.4), the ILMT turns out to be an excellent asteroid hunter.

%The whole document is formatted with the Times New Roman font and a line spacing of 1.15 (double line spacing with the ``manuscript'' option).
%The font size is 12 everywhere except in some titles.
%Text paragraphs are justified to the left and right margins.
%The \textsf{bullsrsl} \LaTeX\ class takes care of all these requirements; it is sufficient to use the adequate sectioning commands.
\end{abstract}


%%\section{Section -- Level 1 title (Times New Roman, bold, 14 pts)}
%\section{\textbackslash section\{\dots\}: Level 1 title}
%The publisher will add
%(i) a header with the bibliographic reference including the Digital Object Identifier (DOI),
%(ii) page numbers
%(iii) a line with the dates of submission and acceptance and
%(iv) the mention of the open access Creative Commons CC~BY~4.0 license.

\section{Introduction}

First light has been obtained with the 4-m ILMT on 29 April 2022 (\citealt{paper15}). The ILMT consists of a high precision photometric and astrometric survey instrument observing at the zenith in the Time Delay Integration (TDI)  mode (see the ILMT poster paper by Surdej et al. for more details). The singly scanned CCD frames correspond to an integration time of 102 sec, i.e. the time an object$^\prime$s image remains within the active area of the detector.
A unique niche for the ILMT is the detection of optical transients for which rapid spectroscopic follow-up observations with the 3.6-m DOT or direct imaging with the 1.3-m DFOT or 3.6-m DOT telescopes can be easily carried out. 

We were thus very much interested in searching for optical transients in the ILMT observations collected during 9 nights in October-November 2022. They consist of 3 times 3 nights using the $g$, $r$ and $i$ Sloan spectral filters. Among these, only the $i$-band observations were obtained during 3 consecutive nights (28-30 October 2022). Our first approach for detecting optical transients has been to start with a visual search of these frames for transients (see Fig.\,\ref{fig1}). These detections were then used to calibrate an automatic search for optical transients making use of the image subtraction technique. We then noticed that optical transients associated with many triplets showed similar angular separations and orientations. They naturally consisted of good asteroid candidates. This was confirmed after consulting the asteroid database of the Minor Planet Center. We then made use of the optical transients detected in the other spectral bands ($g$ and $r$) to extend our search for asteroids and other transient candidates. 
We present here the observations and a detailed analysis of just one of the 20 fields imaged with the ILMT in October--November 2022.

\section{Observations and Detailed Analysis}

\begin{figure}
\centering
\includegraphics[width=0.6\textwidth]{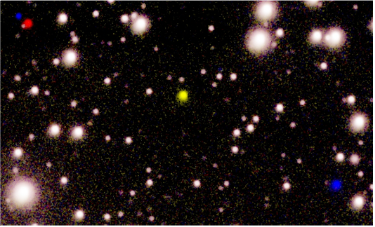}
\bigskip
\begin{minipage}{12cm}
\caption{Excerpt of a RGB-like composite picture consisting of the superposition of three i-band CCD frames recorded with the ILMT on the nights of 28 (blue dot), 29 (yellow dot) and 30 (red dot) October 2022. Near the red dot, an additional blue one is visible corresponding to another optical transient detected on 28 October. Most of the stars which are present on the three frames appear white-like on this picture. Visual inspection of such frames easily leads to the identification of optical transients.
\label{fig1}
}
\end{minipage}
\end{figure}

Each of the fields observed with the ILMT in October-November 2022 covers a rectangular solid angle of $22^\prime \times  198^\prime$ with its length aligned along the right ascension axis. They have been observed during three consecutive nights with the $i$ filter (i.e. the nights of 28, 29 and 30 October 2022). Defining an optical transient as being a source appearing clearly on only one of the three $i$-band frames and an asteroid candidate as consisting of three transients detected at the three different epochs, showing similar angular separations while being properly oriented as a function of time, a total of more than 173 asteroid candidates have been visually identified. Furthermore, we report in these same fields $\sim$ 600 additional $i$-band detections of transients which could be other known or new asteroids, highly variable stars, multiply imaged quasars, supernova candidates, etc. 
Concentrating on just one of those 20 fields, i.e. the field with its RA (2022.8) starting at 04h32m, we have visually identified 22 known asteroids reported in the database of the Minor Planet Center. Making use of the automatic identification of transients based upon the image subtraction technique applied to the three $g$, $r$ and $i$-band CCD frames covering this same field (see \citealt{PranshuKProceedings}), we could identify 53 additional known asteroids reported in the Minor Planet Center database. Furthermore, 48 additional transients have been identified on the basis of the $g$, $r$ and $i$-band CCD frames covering this unique field.
Fig.\,\ref{fig1} illustrates the optical identification of three positions of a same asteroid on the nights of 28 (blue colour), 29 (yellow colour) and 30 October 2022 (red colour).

%\centering
%\begin{minipage}{8cm}
%\caption{Excerpt of a RGB-like composite picture consisting of the superposition of three i-band CCD frames recorded with the ILMT on the nights of 28 (blue dot), 29 (yellow dot) and 30 (red dot) October 2022. Near the red dot, an additional blue one is visible corresponding to another optical transient detected on 28 October. Most of the stars which are present on the three frames appear white-like on this picture. Visual inspection of such frames easily leads to the identification of optical transients.\label{Fig_11}}
%\end{minipage}
%\hfill
%\begin{minipage}{7.5cm}
%\includegraphics[width=7.5cm]{Fig_11.png}
%\end{minipage}
%\end{figure} 

\section{Results and conclusions}

\begin{figure}
\centering
\includegraphics{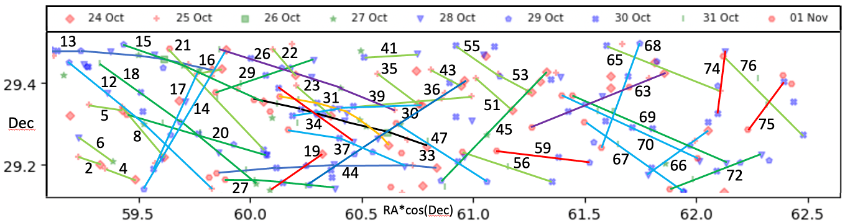}
\bigskip
\begin{minipage}{12cm}
\caption{Paths of the 78 asteroids detected with the ILMT in the 04h 32m LST field during 9 consecutive nights in October-November 2022. The horizontal and vertical axes represent respectively RA cos(Dec) and Dec in degree, where RA is the right ascension and Dec the declination of the asteroids for the 2000 epoch. The numbers refer to the asteroids listed in Table 1. The different coloured symbols correspond to the asteroid positions observed during the different nights.
\label{fig2}
}
\end{minipage}
\end{figure}

\begin{figure}
\centering
\includegraphics[width=14cm]{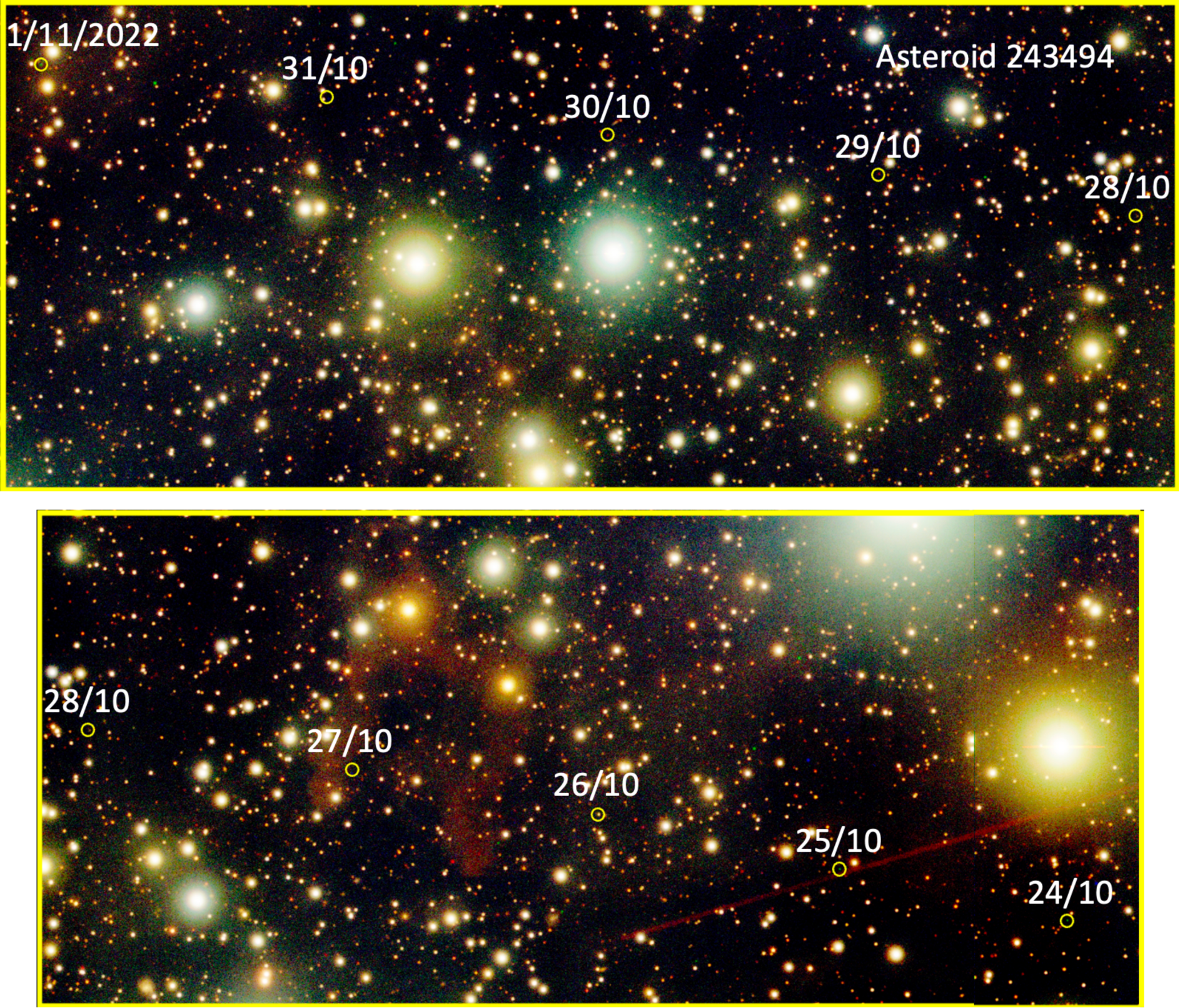}
\bigskip
\begin{minipage}{12cm}
\caption{Example of an asteroid (243494) observed with the ILMT on 9 consecutive nights in October/November 2022 using the $g$, $r$ and $i$ Sloan filters. The broad red streak is the signature of a passing space debris.
\label{fig3}
}
\end{minipage}
\end{figure}

Concentrating on the $22^\prime \times 198^\prime$ field taken at the Local Sidereal Time (LST) 4h 22m, we report a total of 231 optical transient candidates. After consulting the Minor Planet Center database of asteroids, we have found that 214 of these optical transients correspond to the positions of 78 known asteroids brighter than V=22 mag. 
Among these, we find that 1 (resp. 2, 2, 3, 7, 11, 5, 21, 26) asteroids have been detected on 9 (resp. 8, 7, 6, 5, 4, 3, 2, 1) nights (see Table\,\ref{tab1} and Figs.\,\ref{fig2} \& \ref{fig3}). 
Analysis of the remaining CCD frames covering 19 more fields (out of 20) should lead to an impressive number of asteroids observed in just 9 nights. The ILMT turns out to be an excellent asteroid hunter.

\begin{table}
\centering
\begin{minipage}{153mm}
\caption{List of the 78 asteroids detected with the ILMT during 9 consecutive nights in October-November 2022. The symbol \#
indicates their label displayed in Fig.\,\ref{fig2}. The last column indicates the number of nights they were observed.}
\label{tab1}
\end{minipage}
\bigskip

\begin{tabular}{ccccccccc}
\hline
\textbf{\#} & \textbf{Asteroid} & \textbf{Nights} & \textbf{\#} & \textbf{Asteroid} & \textbf{Nights} & \textbf{\#} & \textbf{Asteroid} & \textbf{Nights}\\
\hline
1 & 287600 & 1 & 27 & 285615 & 5 & 53 & 251579 & 2 \\
2 & 398791 & 2 & 28 & 351463 & 8 & 54 & 290252 & 2 \\
3 & 583720 & 1 & 29 & 449883 & 5 & 55 & 568319 & 2 \\
4 & 72368 & 2 & 30 & 475809 & 4 & 56 & 66305 & 2 \\
5 & 7377 & 2 & 31 & 2011 EK46 & 8 & 57 & 2005 UB252 & 1 \\
6 & 121771 & 2 & 32 & 585743 & 2 & 58 & 2016 EL276 & 1 \\
7 & 2009 UD159 & 1 & 33 & 243494 & 9 & 59 & 290365 & 4 \\
8 & 528498 & 2 & 34 & 488998 & 3 & 60 & 599255 & 1 \\ 
9 & 574664 & 1 & 35 & 54958 & 2 & 61 & 2013 HS101 & 1 \\ 
10 & 271811 & 1 & 36 & 59790 & 7 & 62 & 349593 & 1 \\
11 & 2022 SX245 & 1 & 37 & 232625 & 4 & 63 & 356264 & 6 \\
12 & 38733 & 4 & 38 & 2005 UY125 & 1 & 64 & 65781 & 2 \\
13 & 147247 & 7 & 39 & 350968 & 2 & 65 & 351688 & 2 \\
14 & 2011 WM162 & 4 & 40 & 140914 & 1 & 66 & 101033 & 4 \\
15 & 25526 & 5 & 41 & 2022 UQ89 & 2 & 67 & 224216 & 4 \\
16 & 2017 QL25 & 2 & 42 & 270512 & 1 & 68 & 222288 & 5 \\
17 & 87719 & 4 & 43 & 39161 & 3 & 69 & 166170 & 4 \\
18 & 595633 & 6 & 44 & 354803 & 1 & 70 & 89412 & 4 \\
19 & 261417 & 3 & 45 & 91818 & 5 & 71 & 402103 & 1 \\
20 & 273964 & 5 & 46 & 594949 & 1 & 72 & 398828 & 5 \\
21 & 423552 & 2 & 47 & 98222 & 4 & 73 & 2022 UM108 & 1 \\
22 & 130440 & 2 & 48 & 568302 & 1 & 74 & 2001 YP31 & 3 \\
23 & 149624 & 2 & 49 & 2005 UF171 & 1 & 75 & 125328 & 3 \\
24 & 2014 EJ64 & 1 & 50 & 2009 UC166 & 1 & 76 & 250774 & 2 \\
25 & 2019 NX78 & 1 & 51 & 304478 & 2 & 77 & 363975 & 1 \\
26 & 37765 & 6 & 52 & 169617 & 1 & 78 & 568211 & 1 \\
\hline
\end{tabular}
\end{table}

\begin{acknowledgments}
The 4-m International Liquid Mirror Telescope (ILMT) project results from a collaboration between the Institute of Astrophysics and Geophysics (University of Li\`{e}ge, Belgium), the Universities of British Columbia, Laval, Montreal, Toronto, Victoria and York University, and Aryabhatta Research Institute of Observational SciencES (ARIES, India). The authors thank Hitesh Kumar, Himanshu Rawat, Khushal Singh and other observing staff for their assistance at the 4-m ILMT.  The team acknowledges the contributions of ARIES's past and present scientific, engineering and administrative members in the realisation of the ILMT project. JS wishes to thank Service Public Wallonie, F.R.S.-FNRS (Belgium) and the University of Li\`{e}ge, Belgium for funding the construction of the ILMT. PH acknowledges financial support from the Natural Sciences and Engineering Research Council of Canada, RGPIN-2019-04369. PH and JS thank ARIES for hospitality during their visits to Devasthal. B.A. acknowledges the Council of Scientific $\&$ Industrial Research (CSIR) fellowship award (09/948(0005)/2020-EMR-I) for this work. M.D. acknowledges Innovation in Science Pursuit for Inspired Research (INSPIRE) fellowship award (DST/INSPIRE Fellowship/2020/IF200251) for this work. T.A. thanks Ministry of Higher Education, Science and Innovations of Uzbekistan (grant FZ-20200929344). This work is supported by the Belgo-Indian Network for Astronomy and astrophysics (BINA), approved by the International Division, Department of Science and Technology (DST, Govt. of India; DST/INT/BELG/P-09/2017) and the Belgian Federal Science Policy Office (BELSPO, Govt. of Belgium; BL/33/IN12).
\end{acknowledgments}

\begin{furtherinformation}

\begin{orcids}
\orcid{0000-0002-7005-1976}{Jean}{Surdej} 
\end{orcids}

\begin{authorcontributions}
This work results from a long-term collaboration to which all authors have made significant contributions.

%This section is mandatory when there is more than one author.
%The contributions of each author (identified by their initials) must be declared.
%We recommend to follow the \href{http://credit.niso.org}{CRediT} taxonomy (Contributor Roles Taxonomy).
\end{authorcontributions}

\begin{conflictsofinterest}
The authors declare no conflict of interest.
%This section is \emph{mandatory}.
%Authors must declare any personal or professional circumstances that may be perceived as influencing the research reported in the paper.
%If there is no conflict of interest, please state that ``The authors declare no conflict of interest.''
\end{conflictsofinterest}

\end{furtherinformation}

\bibliographystyle{bullsrsl-en}

\bibliography{S11-P18_PospieszalskaSurdejA}

\end{document}